\documentclass[a4paper,12pt, epsfig]{article}
\usepackage{epsfig}
\usepackage{amssymb,latexsym}
\usepackage{amsfonts}
\usepackage{amsmath}
\usepackage{graphicx}

\newskip\humongous \humongous=0pt plus 1000pt minus 1000pt

\newif\ifdtup

\catcode`\@=11

\@addtoreset{equation}{section}
\def\theequation{\thesection.\arabic{equation}}

\def\@normalsize{\@setsize\normalsize{15pt}\xiipt\@xiipt
\abovedisplayskip 14pt plus3pt minus3pt%
\belowdisplayskip \abovedisplayskip
\abovedisplayshortskip \z@ plus3pt%
\belowdisplayshortskip 7pt plus3.5pt minus0pt}

\def\small{\@setsize\small{13.6pt}\xipt\@xipt
\abovedisplayskip 13pt plus3pt minus3pt%
\belowdisplayskip \abovedisplayskip
\abovedisplayshortskip \z@ plus3pt%
\belowdisplayshortskip 7pt plus3.5pt minus0pt
\def\@listi{\parsep 4.5pt plus 2pt minus 1pt
      \itemsep \parsep
      \topsep 9pt plus 3pt minus 3pt}}

\relax

\catcode`@=12

\catcode`\@=11

\def\section{\@startsection{section}{1}{\z@}{3.5ex plus 1ex minus
    .2ex}{2.3ex plus .2ex}{\large\bf}}

\def\thesection{\arabic{section}}
\def\thesubsection{\arabic{section}.\arabic{subsection}}

\def\appendix{\setcounter{section}{0}
  \def\thesection{Appendix \Alph{section}}
  \def\thesubsection{\Alph{section}.\arabic{subsection}}
  \def\theequation{\Alph{section}.\arabic{equation}}}

\def\SymBoxes#1#2#3#4{\newdimen\un@t \un@t#3%
\raisebox{#1}{\rule{#2\un@t}{#4}\hskip-#2\un@t
\@tempdimb\un@t \advance\@tempdimb by-#4\@tempcntb#2\relax%
\@whilenum{\@tempcntb>0}\do{
\rule{#4}{\un@t}\hskip\@tempdimb \advance\@tempcntb by\m@ne}%
\hskip-#2\un@t \rule[\un@t]{#2\un@t}{#4}%
\rule[\un@t]{#4}{#4}\hskip-#4
\rule{#4}{\un@t}}\hskip-#4}                

\begin{document}


\newcommand{\dd}{\textrm{d}}

\newcommand{\beq}{\begin{equation}}
\newcommand{\eeq}{\end{equation}}
\newcommand{\bea}{\begin{eqnarray}}
\newcommand{\eea}{\end{eqnarray}}
\newcommand{\beas}{\begin{eqnarray*}}
\newcommand{\eeas}{\end{eqnarray*}}
\newcommand{\defi}{\stackrel{\rm def}{=}}
\newcommand{\non}{\nonumber}
\newcommand{\bquo}{\begin{quote}}
\newcommand{\enqu}{\end{quote}}
\def\de{\partial}
\def\Om{\ensuremath{\Omega}}
\def\Tr{ \hbox{\rm Tr}}
\def\rc{ \hbox{$r_{\rm c}$}}
\def\H{ \hbox{\rm H}}
\def\HE{ \hbox{$\rm H^{even}$}}
\def\HO{ \hbox{$\rm H^{odd}$}}
\def\HEO{ \hbox{$\rm H^{even/odd}$}}
\def\HOE{ \hbox{$\rm H^{odd/even}$}}
\def\HHO{ \hbox{$\rm H_H^{odd}$}}
\def\HHEO{ \hbox{$\rm H_H^{even/odd}$}}
\def\HHOE{ \hbox{$\rm H_H^{odd/even}$}}
\def\K{ \hbox{\rm K}}
\def\Im{ \hbox{\rm Im}}
\def\Ker{ \hbox{\rm Ker}}
\def\const{\hbox {\rm const.}}
\def\o{\over}
\def\im{\hbox{\rm Im}}
\def\re{\hbox{\rm Re}}
\def\bra{\langle}\def\ket{\rangle}
\def\Arg{\hbox {\rm Arg}}
\def\exo{\hbox {\rm exp}}
\def\diag{\hbox{\rm diag}}
\def\longvert{{\rule[-2mm]{0.1mm}{7mm}}\,}
\def\a{\alpha}
\def\b{\beta}
\def\e{\epsilon}
\def\l{\lambda}
\def\ol{{\overline{\lambda}}}
\def\ochi{{\overline{\chi}}}
\def\th{\theta}
\def\s{\sigma}
\def\oth{\overline{\theta}}
\def\ad{{\dot{\alpha}}}
\def\bd{{\dot{\beta}}}
\def\oD{\overline{D}}
\def\opsi{\overline{\psi}}
\def\dag{{}^{\dagger}}
\def\tq{{\widetilde q}}
\def\L{{\mathcal{L}}}
\def\p{{}^{\prime}}
\def\W{W}
\def\N{{\cal N}}
\def\hsp{,\hspace{.7cm}}
\def\bo{\ensuremath{\hat{b}_1}}
\def\bfo{\ensuremath{\hat{b}_4}}
\def\co{\ensuremath{\hat{c}_1}}
\def\cfo{\ensuremath{\hat{c}_4}}
\newcommand{\C}{\ensuremath{\mathbb C}}
\newcommand{\Z}{\ensuremath{\mathbb Z}}
\newcommand{\R}{\ensuremath{\mathbb R}}
\newcommand{\rp}{\ensuremath{\mathbb {RP}}}
\newcommand{\cp}{\ensuremath{\mathbb {CP}}}
\newcommand{\vac}{\ensuremath{|0\rangle}}
\newcommand{\vact}{\ensuremath{|00\rangle}                    }
\newcommand{\oc}{\ensuremath{\overline{c}}}
\def\P{\ensuremath{\mathcal{P}}}

\newcommand{\Vol}{\textrm{Vol}}

\newcommand{\half}{\frac{1}{2}}

\def\changed#1{{\bf #1}}

\begin{titlepage}
\bigskip
\def\thefootnote{\fnsymbol{footnote}}

\begin{center}
{\Large {\bf
Slow Burgers Vortices in Hot Conformal Fluids
  } }
\end{center}

\bigskip
\begin{center}
{\large  Jarah EVSLIN\footnote{\texttt{jarah@ihep.ac.cn}}}
\end{center}

\renewcommand{\thefootnote}{\arabic{footnote}}

\begin{center}

Institute of High Energy Physics\\
Chinese Academy of Sciences, P.O. Box 918-4, Beijing 100049, P. R. China

\end{center}

\vspace{1.6cm}

\noindent
\begin{center} {\bf Abstract} \end{center}

\noindent
The quintessential vortex solution in (3+1)-dimensional nonrelativistic, incompressible fluid mechanics is the Burgers vortex.  We show that, in a finite domain, conformal fluids also admit hot vortex solutions with everywhere nonrelativistic speeds.  These are identical to Burgers' solution, except that their radius is reduced by a factor of $2/\sqrt{3}$.    A rough calculation indicates that at RHIC these vortices are indeed smaller than the fireball itself during thermalization.   Similarly to the Burgers vortex, these solutions manifest vortex stretching which avoids short distance singularities and so suggests that conformal fluid flows share the same nonsingularity as solutions of the Navier-Stokes equations.  Naively generalizing this calculation to an arbitrary equation of state $w$, we observe that the Burgers vortex radius diverges as $w$ crosses $-1$.  While it has been argued that such a crossover leads to an instability in certain perfect fluids, the absence of Burgers vortices and therefore vortex stretching suggests that, in addition to the well-studied big rip singularities, viscous phantom fluids generically develop vorticity singularities.

\vfill

\begin{flushleft}
{\today}
\end{flushleft}
\end{titlepage}

\hfill{}

\setcounter{footnote}{0}

\section{Motivation}

The initial conditions under which the evolution of the Navier-Stokes equations of fluid dynamics lead to a nonsingular solution are unknown.  However, as has been argued in Refs.~\cite{MKO,Frisch}, numerical simulations indicate that such singularities are avoided by a mechanism called vortex stretching.  In particular, in contrast with Kolmogorov's scale-invariant model of turbulence \cite{Kolmogorov}, turbulent configurations contain random distributions of thin vortices \cite{Townsend}.  These vortices are all deformations of an exact solution of (3+1)-dimensional incompressible hydrodynamics, the Burgers vortex \cite{Burgers}.  As we will review below, the Burgers vortex constantly stretches, pumping vorticity out of potentially singular regions and thus cutting off velocity gradients, ultimately preserving the continuity of the flow.

In this note we will extend Burgers' incompressible fluid solution to the case of a (3+1)-dimensional conformal fluid.  We will demonstrate that, to leading order in an expansion in the inverse temperature, hot vortex solutions with nonrelativistic velocities exist, but with a radius which is smaller than that of the Burgers vortex by a factor of $2/\sqrt{3}$.  In the case of incompressible fluids, the fact that the constant stretching of Burgers vortices yields a nonsingular evolution of the flow has not been proven, but has been supported by numerous numerical simulations.  In the conformal case, even such weak support for nonsingular evolution is absent.  The statistical mechanics of webs of thin vortices is highly dependent upon the theory of their nonaxially-symmetric perturbations \cite{MKO} which will not be discussed in this note, and so may in principle be quite different in the conformal case.  However the fact that such similar solutions also exist and indeed stretch in the conformal case nonetheless implies that they do indeed carry vorticity out of high-vorticity regions, and so may also lead to the nonsingularity of conformal fluid flows.

Such conformal fluids are of interest for a number of reasons.  One reason is that, at temperatures far above the deconfinement transition, the quark gluon plasma is a conformal fluid.  It is a charged conformal fluid with various electromagnetic and color fields, however in heavy ion collision data analysis they are often \cite{Luzum} approximated by uncharged, conformal fluids\footnote{More precisely, the baryon number density does not appear in the equations for the velocity and the thermodynamic variables, and so baryon number conservation may be imposed separately, while the effects of electric charge conservation are negligible \cite{Kolb}.}.  Near the deconfinement transition the conformal anomaly leads to a large bulk viscosity which destroys the conformal approximation, and so this approximation may be trusted at best during the initial evolution of man-made and cosmological quark gluon plasma.  The rapid thermalization of the quark gluon plasma at heavy ion collisions suggests a period of turbulence, and the references above suggest that, at least in nonrelativistic (3+1)-dimensional turbulent flows, Burgers vortices play a crucial role in the redistribution of vorticity.  This thermalization occurs when the quark gluon plasma is much hotter than the deconfinement transition temperature, and so the conformal approximation can be trusted.  

Not only does quark gluon plasma appear on Long Island and near Geneva, but also our universe was once made of quark gluon plasma.  Both heavy ion and cosmological data so far are consistent with the perfect fluid approximation, whereas viscosity plays a crucial role in the Burgers vortex.   However the nature of the turbulent flow plays a key role in understanding the evolution of the initial data of heavy ion collisions to the quark-gluon plasma phase, which is perhaps the greatest roadblock to data analysis.  Thus one may hope that data on the structure of the vorticities is not entirely out of reach.  

Another reason for interest in such conformal fluids is that it has recently been demonstrated \cite{Minwalla} that there is a one-to-one correspondence between their flows and hot black brane solutions to Einstein gravity with a negative cosmological constant.  The vortex solution that will be found below lies within the regime of validity of this correspondence, and so is guaranteed to provide a new solution to Einstein gravity \cite{Chethan2}.  However, as we will consider the fluid flow only in a finite region of spacetime, the gravity dual will not be geodesically complete.

We begin in Sec.~\ref{revsez} with a review of the Burgers vortex solution of the Navier-Stokes equations for an incompressible fluid.  In Sec.~\ref{csez} we generalize this result to hot conformal fluids, solving the relativistic Navier-Stokes equations to leading order in a $v/c$ expansion.  We find that indeed there exist, in a local domain, solutions analogous to the Burgers vortex with a velocity that is everywhere nonrelativistic.  These solutions are characterized by their vorticity profiles, which we see are identical to those of the Burgers vortex but spatially reduced by a factor of $2/\sqrt{3}$.  We also determine the spatial profiles of the various thermodynamic quantities, but stress that these may receive large corrections from relativistic effects.  Finally Sec.~\ref{ultsez} contains more speculative topics.  Ignoring the background expansion, we use crude estimates of parameters of the RHIC fireball to determine the size of a potential Burgers vortex.  This rough estimate suggests that such a vortex indeed would fit within the fireball, however it represents only a slight improvement upon a straight dimensional analysis.  We also generalize this construction to an arbitrary equation of state, finding that phantom fluids do not possess Burgers vortices, and so cannot benefit from the desingularization of vorticity caused by vortex stretching.

\section{Review of Burgers' Solution} \label{revsez}

We will start by reviewing the Burgers vortex solutions to the Navier-Stokes equations for a nonrelativistic, incompressible fluid.  We will be interested in time-independent solutions in 3 spatial dimensions, $x$, $y$ and $z$, with a divergenceless background constant strain field.  The solutions will have an axial symmetry which rotates the $x-y$ plane, and the strain field will point inward along that plane with a magnitude $g$.  In other words, there will be background flow which compresses the fluid along the $x-y$ plane, and extends it in the $z$ direction, with a velocity that scales linearly with distance.  The flow is homogeneous in the sense that it is translation invariant up to a Galilean transformation.  Clearly such a condition may not be imposed upon a relativistic system, as the corresponding Lorentz transformations would commute only up to a Thomas precession term, and so a strain field at relativistic speeds may at most be homogeneous either in the $x-y$ plane or along the $z$ direction.  However, in the present note, we will restrict our attention to flows in which the velocity is always nonrelativistic.  A generalization to the relativistic case should be straightforward, mirroring the analysis of the (2+1)-dimensional case in Ref.~\cite{Chethan}. 

The fact that the velocity varies linearly with respect to the distance in a constant strain field implies that at sufficient large distances, the velocity nonetheless becomes relativistic.  Therefore we will restrict our attention to a bounded spatial region.  Physically this is a reasonable restriction when studying turbulent flows, because the strain fields are only approximately constant in local domains.   While the individual particle velocities in cosmological and heavy ion quark gluon plasmas are heavily relativistic, it is not at present clear whether the flow velocities are also relativistic.   In addition, both cosmological and man-made quark gluon plasmas expand rapidly.  Our divergence-free velocity approximation is only valid if the shear $g$  locally dominates this overall expansion.  For example, in the case of the RHIC plasma, this condition is $g>>10^{23}/sec$  \cite{Schnedermann}.  In fact, without such a large shear, the vortex radius that we will find would barely fit inside of the quark gluon plasma fireball.

The axially-symmetric Burgers vortex solution consists of a superposition of the strain field, described in the previous paragraph, with a rotational flow with angular velocity $f$, which is a function of the cylindrical radial coordinate $\rho=\sqrt{x^2+y^2}.$  This function breaks the homogeneity of the flow, leaving a $U(1)$ rotational symmetry as well as translation symmetries in the $z$ and time directions.  The total velocity ${\bf{v}}(\rho)$ is the sum of the strain field and the rotational flow
\beq
v_x(\rho)=-xg-yf(\rho)\hsp
v_y(\rho)=xf(\rho)-yg\hsp
v_z(\rho)=2gz. \label{Ansatz}
\eeq
The time-independent Navier-Stokes equation consists of 3 components indexed by the coordinate label $k$
\beq
0=v^j\partial_j v^k+\partial^k \P(\rho,z)-\nu\partial^j\partial_j v^k \label{ns}
\eeq
where $\P(\rho,z)$ is the pressure per unit density and $\nu$ is the kinematic viscosity, which is taken to be constant as in this section we are interested in Newtonian fluids.  

The cylindrical symmetry implies that it is sufficient to consider only the $x$ and $z$ components of the Navier-Stokes equations (\ref{ns}).  The $z$ component is quite simple, as the linearity $v_z$ implies that its' second derivatives vanish.  It is simply
\beq
\partial_z \P(\rho,z) = -v^z\partial_z v_z=-4g^2z
\eeq
which is solved by
\beq
\P(\rho,z)=\P_0(\rho)-2g^2z^2. \label{Pnonrel}
\eeq
In other words, the pressure does not respect the $z$ translational symmetry of the system.  Total shifts of the pressure are irrelevant in incompressible fluids, however this implies that in the relativistic case the $z$-translational symmetry will necessarily be broken unless the $\rho$-dependence of $v_z$ and $v_\rho$ can be chosen so that this decrease in pressure is precisely compensated by a relativistic $\gamma$ factor.  If there is no such compensation then, in the case of relativistic conformal fluids, the kinematic viscosity will be minimized in the middle of the vortex, and so the Burgers vortex radius will have a minimum.  

The only remaining component of the Navier-Stokes equation (\ref{ns}) is the $x$ component, which is
\beq
0=x[-f(\rho)^2+g^2+\P_0'(\rho)/\rho]+y[2gf(\rho)+\rho(g+3\nu/\rho^2)f\p(\rho)+\nu f^{''}(\rho)]. \label{nsx}
\eeq
For each value of $\rho$, this must hold everywhere on the corresponding circle of the $x-y$ plane, and so both quantities in the square brackets must vanish separately.  This leads to two equations.  The first equation may be integrated to yield the pressure $\P_0(\rho)$ as a function of a constant of integration $\P_{00}$, the angular velocity $f(\rho)$ and the strain $g$
\beq
\P_0(\rho)=\P_{00}+\int_0^\rho d\rho\p\rho\p(f^2(\rho\p)-g^2). \label{prhononrel}
\eeq
In particular, at large $\rho$ we will see that the angular velocity tends to zero, and so the pressure per density tends to $\P_{00}-g^2\rho^2/2$.  The nonrelativistic approximation is violated before the pressure per density difference is of order one, however the pressure per density is very small in a nonrelativistic, incompressible liquid.  Thus one may expect this approximation to break down at radii such that the velocity is of order $\mathcal{O}(\sqrt{\P_{00}})$, well below the speed of light.

The vanishing of the second square bracket in (\ref{nsx}) may be reexpressed in terms of the vorticity
\beq
\omega(\rho)=\frac{1}{\rho}\partial_\rho (\rho^2 f(\rho)). \label{omega}
\eeq
This is easily integrated to yield the Burgers solution \cite{Burgers}
\beq
\omega(\rho)=c e^{-\frac{g\rho^2}{2\nu}} \label{burg}
\eeq
where $c$ is an arbitrary constant of integration.  Note that at large radii $\rho$ the Burgers solution asymptotically tends to a constant strain field.  The vorticity is concentrated within a vortex of radius
\beq
\rho_B=\sqrt{\nu/g}. \label{rnonrel}
\eeq  

\section{Vortices in Conformal Fluids} \label{csez}

In the case of a conformal fluid, even at nonrelativistic velocities, the pressure's contribution to the moment of inertia is no longer negligible.  Thus the system is not well described by the Navier-Stokes equation.  However translation invariance implies that the stress-energy tensor is still conserved.  We will consider fluids with no other conserved charges, and so the Landau frame fluid flow will be entirely characterized by the conservation of the stress tensor \cite{Landau}
\beq
\partial_\mu T^{\mu\nu}=0\hsp T^{\mu\nu}=P(\eta^{\mu\nu}+4u^\mu u^\nu)-\eta \sigma^{\mu\nu}. \label{rns}
\eeq
Here $P$ is the pressure\footnote{$P$ is not the pressure per density $\P$, which in the case of a conformal fluid is equal to $1/3$.}, $u=v\gamma$ is the relativistic 4-velocity, $\eta$ is the dynamic shear viscosity, and $\sigma^{\mu\nu}$ is the shear.  The first term in the stress tensor describes the perfect fluid approximation, while the second incorporates viscosity.  All functional dependencies have been suppressed, but they are analogous to those of the incompressible case, and we have used the conformality of the system to replace the density with three times the pressure.   To generalize the Ansatz (\ref{Ansatz}) to this setting, we will define the constant $g$ and function $f$ by choosing the 4-velocity
\beq
u_x=-xg-yf\hsp u_y=xf-yg\hsp u_z=2gz\hsp u^0=-u_0=\gamma=\sqrt{1+\rho^2(f^2+g^2)+4g^2z^2}. \label{canz}
\eeq
Note the divergence of the 4-velocity is zero and so the shear may be expressed as
\beq
\sigma^{\mu\nu}=P^{\mu\rho}(\partial_\rho u_\sigma+\partial_\sigma u_\rho)P^{\sigma\nu}\hsp P^{\mu\nu}=\eta^{\mu\nu}+u^\mu u^\nu. \label{strain}
\eeq

In this note we will be interested in conformal flows at nonrelativistic speeds.  Therefore we will expand our results in $v/c$ and only keep the leading orders.   There are now 4 equations to consider, the 4 components of Eq.~(\ref{rns}).  The three spatial components are the relativistic generalizations of the corresponding spatial components of Navier-Stokes, while the time component expresses the conservation of energy.  The conservation of energy equation consists of the derivatives of the two terms in Eq.~(\ref{rns}).  Assuming that the temperature and therefore the density and pressure are time-independent, at leading order in the velocity, the derivative of the first term is proportional to the divergence of $u$ and so vanishes.  The derivative of the second term does not vanish, however it contains elements of the strain with one temporal and one spatial index.  Such mixings are suppressed by a factor of the velocity, and therefore this term is of order $\mathcal{O}(v^2/c^2)$ and we will neglect it.  

In a conformal fluid every transport coefficient is equal to a dimensionless constant multiplied by a power of the temperature $T$. We will make the standard assumption that all particle species are in local equilibrium, so that a single temperature is locally well defined.  This appears to be justified in the case quark-gluon plasma \cite{RHIC}.  In particular, conformal fluids are characterized by two dimensionless constants $c_\eta$ and $c_P$ such that
\beq
P=c_p T^4\hsp
\eta=c_\eta T^3. \label{PT}
\eeq
In the case of the quark gluon plasma, RHIC data suggests $c_\eta\lesssim c_p$ \cite{RHIC2}.  We will be interested in a very hot plasma, in which the temperature $T$ is much greater than the scales of the various derivatives $\partial v/v$ and $\partial T/T$.  In particular this implies that the temperature is much greater than the shear $T>>g$.   In this case, the first term in the stress tensor (\ref{rns}) is much larger than the second.  Indeed, without such an assumption, the derivative expansion of the velocity used in the form of the stress tensor (\ref{rns}) would be inapplicable, and the system would not behave like a fluid.  

To leading order in the high temperature, low velocity expansion, the $z$ component of the equation of motion $(\ref{rns})$ may be approximated by the derivative of only the perfect fluid term of the stress tensor.  Imposing time-independence, the conservation of $z$-momentum becomes
\beq
0=\partial_k [P(\eta^{k3}+4u^k u^z)]\sim\partial_z P+4P\partial_k(2gz u^k)=\partial_z P+16g^2zP
\eeq
which is easily integrated to yield
\beq
P=P_0 e^{-8g^2z^2} \label{Pconf}
\eeq
where $P_0$ only depends on $\rho$.  This result is superficially similar to the incompressible result (\ref{Pnonrel}).  However one much recall that in this case the pressure per density $\P$ is constant, and so the density exhibits the same exponential falloff.  Using the relations (\ref{PT}) one may find the $z$-dependence of the temperature $T$ and dynamic shear viscosity $\eta$
\beq
T=T_0 e^{-2g^2z^2} \hsp
\eta=\eta_0 e^{-6g^2z^2} 
\eeq
where $T_0$ and $\eta_0$ only depend on $\rho$, and in fact themselves obey (\ref{PT}) along with $P_0$.     These relations are of limited use because only the constant and conceivably\footnote{More precisely, using the Ansatz (\ref{canz}) the quadratic terms are correct.  However, as will be described in Sec.~\ref{ultsez}, a fully relativistic treatment requires an Ansatz with at least one additional degree of freedom, which has the potential to modify and even eliminate these quadratic corrections.} the quadratic terms in (\ref{Pconf}) are reliable, the others receive relativistic corrections.

Again axial symmetry implies that the $x$ and $y$ equations of motion are equivalent, and so it will suffice to impose the conservation of $x$-momentum.  In this equation the strain tensor (\ref{strain}) will enter at leading order.  Up to cubic order in $f$ and $g$, the spatial components of the strain tensor $\sigma$ defined in Eq.~(\ref{strain}) are simply those of the nonrelativistic strain tensor $\partial u$.  For example
\beq
\sigma^{xx}=-2g-\frac{2xy}{\rho}f\p+\mathcal{O}(v^3)\hsp
\sigma^{xy}=\frac{x^2-y^2}{\rho}f\p+\mathcal{O}(v^3)\hsp
\sigma^{xz}=\mathcal{O}(v^4).
\eeq
Therefore, in the nonrelativistic approximation, the conservation of $x$-momentum is the condition
\beq
0=\partial_x P+4 u^k\partial_k(Pu^x)-\partial_x(\eta \sigma^{xx})-\partial_y(\eta\sigma^{xy}). \label{xeq}
\eeq
The $P\p u^2$ term is subdominant to the $P\p$ term in the nonrelativistic regime, and so may be dropped.  As was the case with the $x$ component of the Navier-Stokes equation, this equation transforms as a vector under the SO(2) rotation symmetry, and so may be decomposed into a quantity multiplied by $x$ and a quantity multiplied by $y$.  Again, these two quantities must vanish independently.

Using (\ref{PT}) to reexpress the viscosity and pressure in terms of the temperature, the vanishing of the $x$ coefficient yields
\beq
0=(4C_pT_0^3+6c_\eta T_0^2 g)T_0\p+4C_p T_0^4 \rho (g^2-f^2)
\eeq
where $T_0\p$ is the derivative of the temperature with respect to the cylindrical radial coordinate $\rho$.  In the high temperature regime $T_0>>g$ this is easily integrated to yield a global solution for the temperature
\beq
T=T_{00} e^{-2g^2z^2+\int d\rho\p\rho\p(f^2-g^2)}. \label{Tconf}
\eeq
As was the case with Eq.~(\ref{Pconf}), this resembles the nonrelativistic result (\ref{prhononrel}) for the pressure per density ratio.  However, the pressure per density ratio $\P$ is now constant.  Eq.~(\ref{Tconf})  implies that at large distances in both the $z$ and $\rho$ directions, the temperature and therefore the viscosity and pressure shrink.  Eventually this invalidates the high temperature approximation, but only beyond the nonrelativistic regime where (\ref{Tconf}) may be trusted.

The final constraint comes from the $y$ coefficient of (\ref{xeq}), which is the local conservation of angular momentum or more precisely the condition that all of the sources of torque must cancel in a stationary configuration.  Using (\ref{PT}), in the nonrelativistic approximation it is
\beq
0=4c_pT^4(2fg+\rho gf\p)+3c_\eta f\p T^3/\rho+c_\eta T^2(f^{''}T+3f\p T\p) \label{angrel}
\eeq
where all derivatives are taken with respect to $\rho$.  This disagrees with its nonrelativistic counterpart (\ref{nsx}) in two respects.  First, the last term, which measures the spatial derivative of the viscosity, is new.  Using the $\rho$-dependence of $T$ in (\ref{Tconf}), one sees that this derivative yields a factor of $\rho(f^2-g^2)$.  This means that the $f^{''} T$ term is larger than the $3f\p T\p$ term by a factor of order the inverse velocity squared, and so the latter may be dropped in this nonrelativistic approximation. 

 The second difference is that the kinematic viscosity $\nu$ has been replaced with
\beq
\frac{c_\eta}{4c_P T}=\frac{3}{4}\nu \label{eos}
\eeq
where we have used the fact that the density of a conformal fluid is three times its pressure to define a kinematic viscosity $\nu$.   However the new factor of $3/4$ in front of the viscosity changes the Burgers solution (\ref{burg}) to
\beq
\omega(\rho)=c e^{-\frac{2g\rho^2}{3\nu}} 
\eeq
and so the Burgers radius has decreased from the incompressible value of $\sqrt{\nu/g}$.  This is the main result of our analysis, the radius of the Burgers vortex in a conformal fluid subjected to a constant strain $g$ is
\beq
\rho_B=\frac{1}{2}\sqrt\frac{3\nu}{g}. \label{rb}
\eeq

\section{RHIC, Relativistic Vortices and Phantoms} \label{ultsez}

Using an initial RHIC dynamic viscosity $\eta$ of order $10/fm^3$, density of order $1\ GeV/fm^3$ and strain of order $10^{23}/sec$ leads to a Burgers radius of $\rho_B\sim .1-1 fm$.  The Burgers vortex radius is inversely proportional to the temperature, and so increases by an order of magnitude prior to freeze-out, eventually being about the same size as the entire fireball and invalidating the solution.   Needless to say, the fact that the RHIC fireball expands means that the strain is not divergenceless, as has been assumed here.  Burgers vortices in an expanding fluid will be treated elsewhere.  Such a treatment is a prerequisite for a determination of the observational consequences (or lack thereof) of these vortices in the quark gluon plasma at the LHC.

The RHIC quark gluon plasma achieves thermal equilibrium, and so one may define a (convention-dependent) local velocity.  It is not known whether this velocity is relativistic, but if it is, then one must confront the inconsistency of our Ansatz (\ref{canz}).  In principle a consistent Ansatz requires that the coefficient functions in the velocity components possess a nontrivial $z$-dependence.  However if the fully relativistic solutions share a $z$-translation symmetry up to a Lorentz boost, generalizing the longitudinal symmetry of Burgers solutions, then the $z$-dependence will be dictated entirely by this symmetry and a constant characterizing the strain.  Our Ansatz (\ref{canz}) is nonetheless inconsistent, as there are four equations of motion but only two free functions $f$ and $T$ of $\rho$.   We have argued that to leading order in the velocity our Ansatz contains solutions, which are justed rescaled Burgers vortices. 

In a fully relativistic treatment the velocity components $u_z$ and $u_\rho$ must also be functions of $\rho$ determined by the equations of motion.  As $u_z$ is already a function of $z$, this naively leads to partial differential equations, which would be quite difficult to solve numerically.  However, if one fixes $u_z$ at $z=0$, then the value of $u_z$ may be determined everywhere by imposing the aforementioned longitudinal translation/boost symmetry.  This leaves four ordinary differential equations at $z=0$ and four undetermined functions of $\rho$: the longitudinal, angular and radial velocities and the temperature at $z=0$.  It therefore seems plausible that such a translation/boost-symmetric vortex exists.  Note however that in this counting we have allowed $u_z$ at $z=0$ to be an arbitrary function of $\rho$.  Therefore it seems likely that this function will necessarily be nonzero, and so the relativistic vortex will not share the $z\rightarrow -z$ reflection symmetry of the incompressible Burgers vortex, instead these vortices would come with orientations. 

What would be the physical interpretation of $u_z(\rho)$ at $z=0$?  Naively it seems like a center of mass flow, however by shifting the coordinate $z$, one can arrange for the net $z$-flow at $z=0$ to vanish, for example leaving $T^{tz}(z=0)$ positive for $\rho<\rho_c$ and negative for $\rho>\rho_c$ where $\rho_c$ depends entirely on the vorticity, background shear and background temperature.  Of course, even if such bizarre solutions exist, it does not mean that they play an important role in the hydrodynamics of these fluids.  The above degree of freedom count merely suggests that such a feature may be necessary if one demands the same longitudinal translation/boost symmetry enjoyed by vortices in incompressible fluids.

One may readily generalize these vortices to a nonconformal fluid with density $e$, no bulk viscosity and an equation of state
\beq
w=\frac{P}{e}.
\eeq
We will not assume that $w$ is a universal constant for a given fluid, as negative values of $w$ would then lead to an imaginary speed of sound and so an instability.  However, motivated by the small fractional change in temperature seen above over the radial profile of the vortex solution, we will assume that $w$ assumes a constant value on the entire vortex solution.  Then, again dropping the subdominant derivative of the viscosity, the conservation of angular momentum (\ref{angrel}) becomes
\beq
0=\left(\frac{w+1}{w}\right) Pg (2f+\rho g\p)+\eta(\frac{3f\p}{\rho}+f^{\prime\prime})
\eeq
and so the Burgers radius is 
\beq
\rho_B(w)=\sqrt{\frac{\nu}{(1+w)g}}.
\eeq
 In the incompressible and conformal cases $w=0$ and $w=1/3$ this yields Eqs.~(\ref{rnonrel}) and (\ref{rb}) as it must.

Not surprisingly, there is no vortex solution for dark energy $w=-1$.  However  Burgers vortices also do not exist for phantom fluids $(w<-1)$, unless the kinetic viscosity $\nu$ is negative.  In fact, without such a negative $\nu$, the conservation of angular momentum equation (\ref{angrel}) implies that the viscosity increases, rather than decreasing, the angular velocity in such configurations and therefore probably leads to a singularity, or at least a breakdown of local equilibrium.   Ignoring such potential instabilities, if $\nu$ remains positive there will be no Burgers vortex solutions and so  the vortex stretching mechanism for avoiding singularities in incompressible fluids does not apply to phantom fluids.  Thus, it seems quite plausible that such fluid flows develop singular vorticities where the local equilibrium conditions and therefore the hydrodynamic description itself break down.

Phantom fluids do not satisfy the null energy condition.  A large class of microscopic theories described by such fluids have been shown to either be unstable or to admit superluminal propagation\footnote{It has been shown \cite{Nima} that in some such theories superluminal propagation leads to closed timelike curves (CTCs) and therefore nonlocal constraints, which renders the Cauchy problem ill-defined.   However in Refs.~\cite{AlexDifesa1,AlexDifesa2} it is suggested that a universal time prevents CTCs in some of the most popular examples.} in Ref.~\cite{0512260}.  In the case of perfect fluids microscopically described by a single scalar field with a reasonably general Lagrangian, obstructions to crossing\footnote{A combination of globular cluster, white dwarf and supernova data appears to favor such a crossover~\cite{quintom,quintomreview}.} $w=-1$  have already been described in Refs.~\cite{quintom,Alex,Zhang}.  However the present argument is generic to any fluid with a positive kinematic shear viscosity and $w<-1$, even if it has never crossed $w=-1$ and regardless of its microscopic degrees of freedom and even if, like the theory in Ref.~\cite{0312099}, it escapes the instability argument of  \cite{0512260}.  However, even if such an instability does exist, it may be avoided if $w<-1$ for a sufficiently limited duration \cite{rip}.



\section* {Acknowledgment}   

\noindent
I would like to thank Chethan Krishnan and Xinmin Zhang for useful discussions.   I am supported by the CAS Fellowship for Young International Scientists grant number 2010Y2JA01.



\begin{thebibliography}{99}

\bibitem{MKO}
H.~K.~Moffatt,  S.~Kida and K.~Ohkitani, ``Stretched vortices - the sinews of turbulence; large Reynolds number asymptotics'', J. Fluid Mech. {\bf{259}}  241-264 (1994).

\bibitem{Frisch}
U. Frisch, ``Turbulence:  The Legacy of A. N. Kolmogorov", Cambridge 1996.

\bibitem{Kolmogorov}
L.~F. Richardson, ``Weather Prediction by Numerical Process.'' Cambridge: Cambridge University Press, 1922.
A.~N.~Kolmogorov, ``The local structure of turbulence in incompressible viscous fluids for very large Reynolds numbers", Dokl. Akad. Nauk. SSSR {\bf{30}} 9-13 (1941)  (reprinted in Proc. R. Soc. Lond. A {\bf{434}}, 9-13 (1991)).
A.~N.~Kolmogorov, ``On degeneration (decay) of isotropic turbulence in an incompressible viscous liquid", Dokl. Akad. Nauk. SSSR {\bf{31}} 538-540 (1941) .
A.~N.~Kolmogorov, ``Dissipation of energy in locally isotropic turbulence",  Dokl. Akad. Nauk. SSSR {\bf{32}} 16-18 (1941) (reprinted in Proc. R. Soc. Lond. A {\bf{434}},  15-17 (1991)).

\bibitem{Townsend}
A.~A.~Townsend, ``On the fine scale structure of turbulence", Proc. R. Soc. Lond. A {\bf{208}} 534-542 (1951) .

\bibitem{Burgers} 
J. M. Burgers, ``A mathematical model illustrating the theory of turbulence", Adv. in Appl. Mech {\bf{1}} 171-199 (1948) .

\bibitem{Luzum}
  M.~Luzum and P.~Romatschke,
  ``Conformal Relativistic Viscous Hydrodynamics: Applications to RHIC results
  at $\sqrt{s_{NN}}$ = 200 GeV,''
  Phys.\ Rev.\  C {\bf 78} (2008) 034915
  [Erratum-ibid.\  C {\bf 79} (2009) 039903]
  [arXiv:0804.4015 [nucl-th]].

\bibitem{Kolb}
  P.~F.~Kolb and U.~W.~Heinz,
  ``Hydrodynamic description of ultrarelativistic heavy-ion collisions,''
  arXiv:nucl-th/0305084.


\bibitem{Minwalla}
  S.~Bhattacharyya, V.~EHubeny, S.~Minwalla and M.~Rangamani,
  ``Nonlinear Fluid Dynamics from Gravity,''
  JHEP {\bf 0802}, 045 (2008).
  [arXiv:0712.2456 [hep-th]].

\bibitem{Chethan2}
  J.~Evslin and C.~Krishnan,
  Work in progress.

\bibitem{Chethan}
  J.~Evslin and C.~Krishnan,
  ``Vortices in (2+1)d Conformal Fluids,''
  JHEP {\bf 1010}, 028 (2010).
  [arXiv:1007.4452 [hep-th]].

\bibitem{Schnedermann}
  E.~Schnedermann and U.~W.~Heinz,
  ``A Hydrodynamical assessment of 200-A/GeV collisions,''
  Phys.\ Rev.\  C {\bf 50} (1994) 1675
  [arXiv:nucl-th/9402018].

\bibitem{Landau}
  L.~D.~Landau,
  ``On the multiparticle production in high-energy collisions,''
  Izv.\ Akad.\ Nauk Ser.\ Fiz.\  {\bf 17} (1953) 51.

\bibitem{RHIC}
  K.~Adcox {\it et al.}  [PHENIX Collaboration],
  ``Formation of dense partonic matter in relativistic nucleus nucleus
  collisions at RHIC: Experimental evaluation by the PHENIX  collaboration,''
  Nucl.\ Phys.\  A {\bf 757} (2005) 184
  [arXiv:nucl-ex/0410003].

\bibitem{RHIC2}
 J.~Adams {\it et al.}  [STAR Collaboration],
  ``Experimental and theoretical challenges in the search for the quark  gluon
  plasma: The STAR collaboration's critical assessment of the  evidence from
  RHIC collisions,''
  Nucl.\ Phys.\  A {\bf 757} (2005) 102
  [arXiv:nucl-ex/0501009].
 
\bibitem{0512260}
  S.~Dubovsky, T.~Gregoire, A.~Nicolis and R.~Rattazzi,
  ``Null energy condition and superluminal propagation,''
  JHEP {\bf 0603} (2006) 025
  [arXiv:hep-th/0512260].

\bibitem{Nima}
  A.~Adams, N.~Arkani-Hamed, S.~Dubovsky, A.~Nicolis, R.~Rattazzi,
  ``Causality, analyticity and an IR obstruction to UV completion,''
  JHEP {\bf 0610}, 014 (2006).
  [hep-th/0602178].

\bibitem{AlexDifesa1}
  E.~Babichev, V.~Mukhanov, A.~Vikman,
  ``k-Essence, superluminal propagation, causality and emergent geometry,''
  JHEP {\bf 0802}, 101 (2008).
  [arXiv:0708.0561 [hep-th]].

\bibitem{AlexDifesa2}
  O.~Pujol{\`a}s, I.~Sawicki and A.~Vikman,
 ``The Imperfect Fluid behind Kinetic Gravity Braiding,"
 [arXiv:1103.5360 [hep-th]].


\bibitem{quintom}
  B.~Feng, X.~L.~Wang and X.~M.~Zhang,
  ``Dark Energy Constraints from the Cosmic Age and Supernova,''
  Phys.\ Lett.\  B {\bf 607} (2005) 35
  [arXiv:astro-ph/0404224].

\bibitem{quintomreview}
  Y.~F.~Cai, E.~N.~Saridakis, M.~R.~Setare and J.~Q.~Xia,
  ``Quintom Cosmology: Theoretical implications and observations,''
  Phys.\ Rept.\  {\bf 493} (2010) 1
  [arXiv:0909.2776 [hep-th]].

\bibitem{Alex}
  A.~Vikman,
  ``Can dark energy evolve to the phantom?''
  Phys.\ Rev.\  D {\bf 71} (2005) 023515
  [arXiv:astro-ph/0407107].

\bibitem{Zhang}
  G.~B.~Zhao, J.~Q.~Xia, M.~Li, B.~Feng and X.~Zhang,
  ``Perturbations of the Quintom Models of Dark Energy and the Effects on
  Observations,''
  Phys.\ Rev.\  D {\bf 72} (2005) 123515
  [arXiv:astro-ph/0507482].

\bibitem{0312099}
  N.~Arkani-Hamed, H.~C.~Cheng, M.~A.~Luty and S.~Mukohyama,
  ``Ghost condensation and a consistent infrared modification of gravity,''
  JHEP {\bf 0405} (2004) 074
  [arXiv:hep-th/0312099].

\bibitem{rip}
  T.~Qiu, Y.~F.~Cai and X.~M.~Zhang,
  ``Null Energy Condition and Dark Energy Models,''
  Mod.\ Phys.\ Lett.\  A {\bf 23} (2008) 2787
  [arXiv:0710.0115 [gr-qc]].





\end{thebibliography}
\end{document}